\documentclass[review]{elsarticle}
\usepackage[american]{babel}
\usepackage{amsmath}
\usepackage{bbold}
\usepackage{lineno,hyperref}

\journal{Journal of \LaTeX\ Templates}

\biboptions{authoryear,sort,comma}

\begin{document}

\begin{frontmatter}

\title{Clustering of check-in sequences using the mixture Markov chain process.}


\author[mymainaddress]{Elena Shmileva\corref{mycorrespondingauthor}}
\cortext[mycorrespondingauthor]{Corresponding author}
\ead{elena.shmileva@gmail.com}

\author[mymainaddress]{Viktor Sarzhan}
\ead{sarzhanva@gmail.com}

\address[mymainaddress]{National Research University Higher School of Economics, St.Petersburg campus, Math department, \ 3A Kantemirovskaya Street, St Petersburg, 194100, Russian Federation}

\begin{abstract}
This work is devoted to the clustering of check-in sequences from a geosocial network. We used the mixture Markov chain process as a mathematical model for time-dependent types of data. For clustering, we adjusted the Expectation-Maximization (EM) algorithm. As a result, we obtained highly detailed communities (clusters) of users of the now defunct  geosocial network, Weeplaces. 
\end{abstract}

\begin{keyword}
Mixture Markov chains process \sep EM algorithm \sep check-in \sep customer behavior \sep customer journey maps, geosocial networks \sep urban data \sep communities detection.

{\it{Declarations of interest: }}none
\end{keyword}

\end{frontmatter}


\section{Introduction}

We leave digital fingerprints by our daily activity, such as phone calls, check-ins, electronic payments in supermarkets or transactions. Each of these activities produce data, by the analysis of which, we could reconstruct a lifestyle of a person. For this behavior reconstruction, the data scientists actively started to analyze sequences of check-ins from location-based social networks (LBSNs) or/and sequences of bank transactions. These two directions have been studied extensively in recent years  \citep[see][]{cjm-ex'17, gonzalez'18,  cjm'17,   wei, liu'13, vaganov'17}. 
In particular, researchers are  clustering  the sequences and predicting the next point-of-interest (POI) of a user.

 Analyzing users’ bank transactions seems to be a more informative study, because it is often supplemented by  additional information about the users. The most remarkable article in this direction is the work by M.Gonzalez et al., \citep[see][]{gonzalez'18}, where the researchers made clustering of sequences of bank transactions in Mexico City. They used a technique of text contraction, so called the Sequitur algorithm, and then constructed a graph from the statistically significant sequences, and then applied the Louvain algorithm for detecting communities in networks. Supplementary information about the users and their mobile phone activities helped the researchers to understand the communities that were discovered. Similarly, Vaganov et al. analyzed the sequences of transactions from a Russian source - "Bank Saint Petersburg", \citep[see][]{vaganov'17}. They used K-means clustering for N-gramm vectors.

There is a series of works on the POI predictions, based on check-in data from Weeplaces, Gowalla, Foursquare and other LBSNs. The approaches that are used in this area of research are quite diverse, including time-aware POI recommendation, geographical influence POI recommendation \citep[see][]{wei}, and content-aware or socially influenced enhanced POI recommendation. The approach which is closer to the subject of our work is the time-aware POI recommendation. Here, the researchers make their analysis mostly by exploiting the temporal structure of the sequences. Usually, the best results are achieved by combining clustering and a POI prediction. We could mention article by X.Lui et al. \citep[see][]{liu'13} as an example of the time-aware POI recommendation in LBSNs. In this work, US and EU cities are considered. The main tools in this article are the K-means clustering and the matrix factorization or the third-rank tensor.

The existing studies have applied many algorithms of  clustering for sequences of activities. Nevertheless, we have not found any results with soft clustering based on Markov chains for sequences of check-ins or bank transactions. The major challenge and the main contribution of our research is to adjust the mathematically-rich mixture Markov chain model for the clustering of check-in sequences. Usage of this tool highlights many similar problems dealing with sequences of temporally-dependent data. 

In this work, we clusterize sequences of check-ins from the Weeplaces LBSN. The number of clusters is given in advance. As a mathematical model for the check-in sequences, we used the mixture Markov chain process. For clustering, we applied  the Expectation-Maximization (EM) algorithm, which produces a soft clustering, i.e. the result of this algorithm is 
a vector of posterior probabilities for clusters (the mixing distribution of the mixture Markov chain process), as well as parameters of all Markov chains. The vector of posterior probabilities for clusters helps to estimate sizes of the clusters. The parameters of the Markov chains help to describe clusters.

 The clustering of data with the mixture Markov chain model was considered in several works  \citep[see][]{cjm'17, cjm-ex'17}, where the authors used data from a special survey on citizens’ activities, made by the Chicago Metropolitan Agency for Planning. The data were obtained from a representative sample of citizens who agreed to participate in the project, which is why the data are well organized and contain a lot of additional information about the citizens. This situation differs from the check-in data analysis, but, academically, our study is close to this research.

The practical interest in this problem could be explained from both sociological and technological sides. From the sociological viewpoint, the clusters collect users with similar behavior. So, this research would help sociologists to understand some attributes of these communities without doing expensive surveys. For example, our study helps to calculate the sizes of communities and to understand typical consumer trajectories of people from the communities, so called customer journey maps (CJM). From the technological point of view, our study helps to forecast the next POI of users. By our method, one can detect the most probable cluster for a given user. Then one can calculate the stationary distribution of the Markov chain corresponding to this cluster. This allows us to predict the most typical trajectory for this user and to make a forecast.

\section{Methods}

\subsection{Notations}
Let us start with  notations.
We denote the set of categories  by
$CAT$ and the number of categories by
$$C=\#CAT.$$

Let $S$ be the set of all possible sequences of check-ins with values in $CAT$ and
$${\cal{N}}=\#S.$$
We prolong any  finite  sequence $s\in S$, putting $0$ at each step after the step of the last check-in. Thus, we have a set of infinite sequences $S\subset{\cal{S}}=\mathbf{N}\times CAT\cup\{0\}$.

We use a soft clustering (a probabilistic clustering). Each cluster will be described by a Markov chain with sample paths in ${\cal{S}}$, i.e. we deal with Markov chains at a discrete time  with a discrete set of states. 
The mathematical model that we will use   is the mixture Markov chains process.

\newdefinition{definition}{Definition}
\begin{definition}
Mixture Markov chains process  $M(t), t\in \mathbf{N}$ is defined as a mixture of $K$ stochastic processes $M_1(t), ..., M_K(t),$ $t\in \mathbf{N}$ that are  the Markov chains. Namely, according to the mixing distribution, $p=\{p_1, ... , p_K\}$  ($0<p_i<1$ s.t. $\sum_{i=1}^K p_i=1$) 
 \begin{equation*}
Law (M)=p_1\cdot Law (M_1)+...+ p_K\cdot Law(M_K).
 \end{equation*}
\end{definition}

In this work, we will deal just with the stationary Markov chains (or time-homogeneous Markov chains), i.e. $$P(M_i(t+1)=x | M_i(t)=y)= P(M_i(t+\ell+1)=x | M_i(t+\ell)=y)$$ for any $\ell\in \mathbf{N}$ and $i= 1..K$.
 
It is known that any stationary Markov chain is defined by the vector of initial probabilities  $f$ and the transition probability matrix $T$.
Thus, the mixture Markov chains process  is defined by the discrete distribution $p$ on $\{1,..., K\}$ and  the parameters of each Markov chain $M_i$:  $$(f_1, T_1), ..., (f_K, T_K),$$
where  $f_i=(f_i^1, ..., f_i^C)\in [0,1]^C,\  i=1..K$  such that $ \sum_{j=1}^C f_i^j=1$ is the vector of initial probabilities of $M_i$, and 
$T_i\in [0,1]^{C\times C}$ is the transition probability matrix of $M_i$ (i.e. it is a stochastic matrix, namely, $\sum_{j=1}^C T_i(k,j)=1$ for each $k=1..C$).

We choose the number of clusters $K$ artificially. In our work, we will have K=4, K=5 or K=6. We left the problem of finding the optimal number of clusters  for further research.

We will also need the likelihood function:
\begin{equation}\label{likelyhood}
L(p,\theta)=\Pi_{s\in S}P_{p, \theta}(M=s),
\end{equation}
here $M$ is a random element of the mixture Markov chain process,  $p$ is the mixing distribution,  $\theta =(\theta_1, ..., \theta_K)$, $\theta_i=(f_i,T_i)$ are the parameters of the Markov chains, $P_{p, \theta}(M=s)$ is the probability that the sample path of the process $M$ will be equal to a concrete sequence $s\in S$.

By the law of  total probability, where the group of jointly exhaustive events is equal to the events that $s$ belongs to any of given $K$ clusters,
$$P_{p, \theta}(M=s)= P(M=s| \theta_1)\cdot p_1+...+P(M=s| \theta_K)\cdot p_K.$$

By  the properties of Markov chains
\begin{equation}\label{chain_rule}
P(M=s \ | \ \theta_i)= P(s | \theta_i)=f_i(s_1)\cdot T_i(s_1,s_2)\cdot ... \cdot T_i(s_{l_s-1}, s_{l_s}),
\end{equation}
here $s=(s_1, ..., s_{l_s})\in S$,  where $l_s$ is the length of the given sequence $s$.


\subsection{Algorithm}
To optimize the parameters of the mixture Markov chain process, we use the EM algorithm.

The initial values:

the  initial probabilities  for the mixing distribution $p_i=1/K,$ the initial parameters of the Markov chains are
$$
f_i= (1/C, ...,1/C),
$$

$$
    T_i=\left( \begin{array}{ccc} 
     1/C & \ldots & 1/C   \\
      \vdots &  \ddots & \vdots    \\
   1/C & \ldots & 1/C \\
  \end{array} \right).
$$
We also tried this algorithm with random initial values. There were no significant differences  in the results.

E-step:

We introduce the vector of posterior probabilities

$$G_s= \left(P( \theta_1 |  s), ..., P( \theta_K | s)\right)= (g_{s,1}, ..., g_{s,K}),$$

i.e. $g_{s,i}$  is the probability that a given sequence $s\in S$ belongs to the i-th cluster.

Mathematically,
\begin{equation}\label{posteriors}
g_{s,i}=\frac{p_i\cdot P(s | \theta_i)}{\sum_{j=1}^K p_j\cdot P(s | \theta_j)},
\end{equation}
where $P(s | \theta_i)$ is calculated by~(\ref{chain_rule}).

M-step:

At this step, we maximize the likelihood function $L(p,\theta)$ from~(\ref{likelyhood}). 
Namely, we will maximize $$\log L(p,\theta)=\sum_{s\in S} \log \left(\sum_{i=1}^K p_i\cdot P(s | \theta_i)\right)\to \max_{p, \theta}.$$

The solution of this optimization problem is given below. 
The prior probabilities for the next step of the algorithm are recalculated by:
$$p_i=\frac{\sum_{s\in S} g_{s,i}}{\cal{N}}.$$

The initial probabilities for the next step of the algorithm are recalculated as follows:
$$f_i= (f_i^1, ..., f_i^C) \cdot \left(\sum_{j=1}^C f_i^j\right)^{-1},$$
here
$$f_i^j= \sum_{s\in S}g_{s,i}\cdot \mathbb{1}_{\{s_1=j\}}, $$
where $j=1..C$, $i=1..K$ and $f_i^j$ counts the number of sequences  $s$ that start from the category $j\in CAT$ multiplied  by the probability that $s$ belongs to the i-th cluster. 

The transition probability matrices
$T_i=(T_i^1, ..., T_i^C)^T$,  $i=1..K$ are calculated by rows. The j-th row of the matrix $T_i$, is obtained as follows:
$$T_i^j=(t_i(j,1), ..., t_i(j,C))\cdot \left(\sum_{k=1}^C t_i(j,k)\right)^{-1},$$
where
$$t_i(j,k)=\sum_{s\in S}g_{s,i}\cdot \sum_{p=2}^{l_s}\mathbb{1}_{\{s_{p-1=j}\}}
\cdot\mathbb{1}_{\{s_p=k\}},$$
here the number of transitions from $j$ to $k$ in the sequence $s$ is weighted by the probability that $s$ belongs to the i-th cluster.

There is a convergence of this procedure to  $max L(p,\theta)$ by the properties of the EM algorithm.
We control the closeness to the maximum by the following stop condition:
 $$\Delta L(p, \theta)<\varepsilon,$$ where $\varepsilon$ is a given small number and $\Delta L(p, \theta)$ is the difference of the values of the likelihood function  $L(p, \theta)$ at  two consecutive steps of the EM algorithm.


\subsection{Data}
The data are taken from the site of Singapore University, yongliu.org. The data contain the users’ check-in history in a location-based social network called Weeplaces. This social network was a prototype of modern Swarm Foursquare, and it does not exist any longer. This dataset contains 7,658,368 check-ins generated by 15,799 users from 971,309 locations. 
The data are arranged as follows: 

{\bf{userid}} - Login of a user;

{\bf{placeid}} - Name of the geotag where the user made the check-in;

{\bf{datatime}} - Time of the check-in; 

{\bf{lat lon}} - Latitude and longitude of the check-in location;

{\bf{city}} - City of the check-in;

{\bf{category}} - Type of the check-in location (Food, Art $\&$ Entertainment, College $\&$ Education, Home/Work, Nightlife, Parks $\&$ Outdoors, Shops, Travel)

This means that $C=8$ and 
\begin{multline*}
CAT= \{Food, \ Art\ \&\ Entertainment, \ College\ \&\ Education, \ Home/Work,\\ Nightlife,\ \ Parks \ \&\ Outdoors,\ 
Shops, \ Travel\}.
\end{multline*}

From the set of available data, we use only: {\bf{userid}}, {\bf{datetime}}, {\bf{city}}, and {\bf{category}}. {\bf{Placeid}} and {\bf{lat lon}}  are not used in this work, but contain additional information for further research.  The data also contain information about friendship between users, but we also didn't use this information yet.

The data were gathered from 2006 to 2011. We only used  the 2009 to 2011 data, because the 2006 to 2008 activity was negligible.
We also ignored the customers with less than 10 check-ins.

We preprocessed the data as described below.
For a given city, we collected the sequences of activities $s=(s_1, s_2, ..., s_k)$, $k\in \mathbf{N}$, $s\in S$ of each user at a certain time interval. 
The time interval is chosen to be one week for the sake of convenience for computations. 


For example, the sequence $s=(3, 2, 1, 0, 0, 0, ...)$ means that a user  made a check-in at a place from the category Home/Work (category 3), then was at a restaurant or a food court (category 2), and at the end of this time interval she/he made a check-in at a college or at a university (category 1). 

Each user could be represented by several sequences, but we average the number of sequences per user. We must do this to escape the bigger influence of the most active users to the formation of clusters.

\section{Results}

We demonstrate the results of our method for London.

In London, we discovered 134,196 activities for 985 users. We rearranged the data in 7,052 sequences.  It is left 5,467 after elimination of the  short sequences.
Each user could be represented by several sequences of activities. For example, "fred-wilson" is represented by 58 sequences, "ben-parr" is represented  only by 2 sequences in London. Because of the difference in their activity, they have various influence for  clustering. We average the number of sequences per user  as follows: we count the median of sequences per user, $Me$, (for London $Me=5$), and we leave for clustering $Me$ randomly chosen sequences for each user (if a user has less than $Me$ sequences, we leave all of them).


We wanted to obtain 4 clusters. By our method based on sequences, we obtained the posterior probabilities vector: $$p^{seqs}=(0.21, 0.09, 0.28, 0.42).$$

This distribution  shows the sizes of the clusters of the check-ins' sequences. 


We could recompute these probabilities in terms of users, taking into account the origin of each sequence:
 $$p^{users}=(0.22, 0.04, 0.26, 0.48).$$
 We did it by averaging.
 Let $S_{u,London}$ be the set of sequences of user $u$ for London, and  $U$ be the set of users,  then
 $$p^{u}_i=\frac{\sum_{s\in  S_{u,London}} g_{s,i}}{\#S_{u,London}},$$
where $g_{s,i}$ are the probabilities that the sequence $s$ belongs to the i-th cluster see (\ref{posteriors}), and  finally
$$p^{users}_i=\frac{\sum_{u\in  U} p_i^u }{\#U}.$$
Obviously, $\sum_{i=1}^K p_i^{users}=1$.

The vectors of initial probabilities for each cluster are given in Figure 1.
\begin{figure}
\centering
\includegraphics[width=0.5\textwidth]{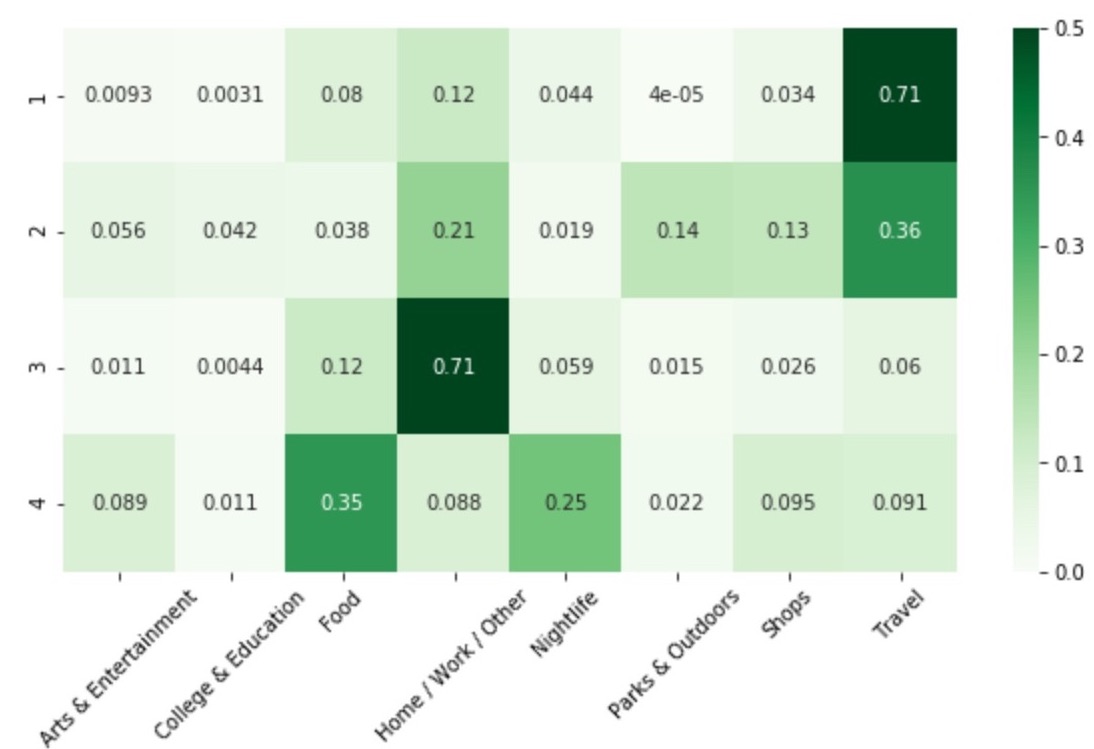}
\caption{London. K=4. Vectors of initial probabilities for each of 4 clusters.}\label{sample image}
\end{figure}

This means that the majority of users in the biggest cluster 4, containing $42\%$ of the sequences (and $48\%$ of users), start their week from the Food or Nightlife check-in. Cluster 3, containing $28\%$ of sequences (and $26\%$ of users) accommodates check-in sequences that start mainly from the Home/Work category.
Cluster 1 contains $21\%$ of sequences (and $22\%$ of users), and the prevalent initial category  in this cluster is Travel. Cluster 2, containing $9\%$ of sequences (and $4\%$ of users) has approximately equal  probabilities to start the week's check-in sequence from any given category (but Travel is a bit more popular).  


 

The  transition probability matrices for each cluster are given in  Figure 2 and Figure 3.
\begin{figure}
\centering
\includegraphics[width=1\textwidth]{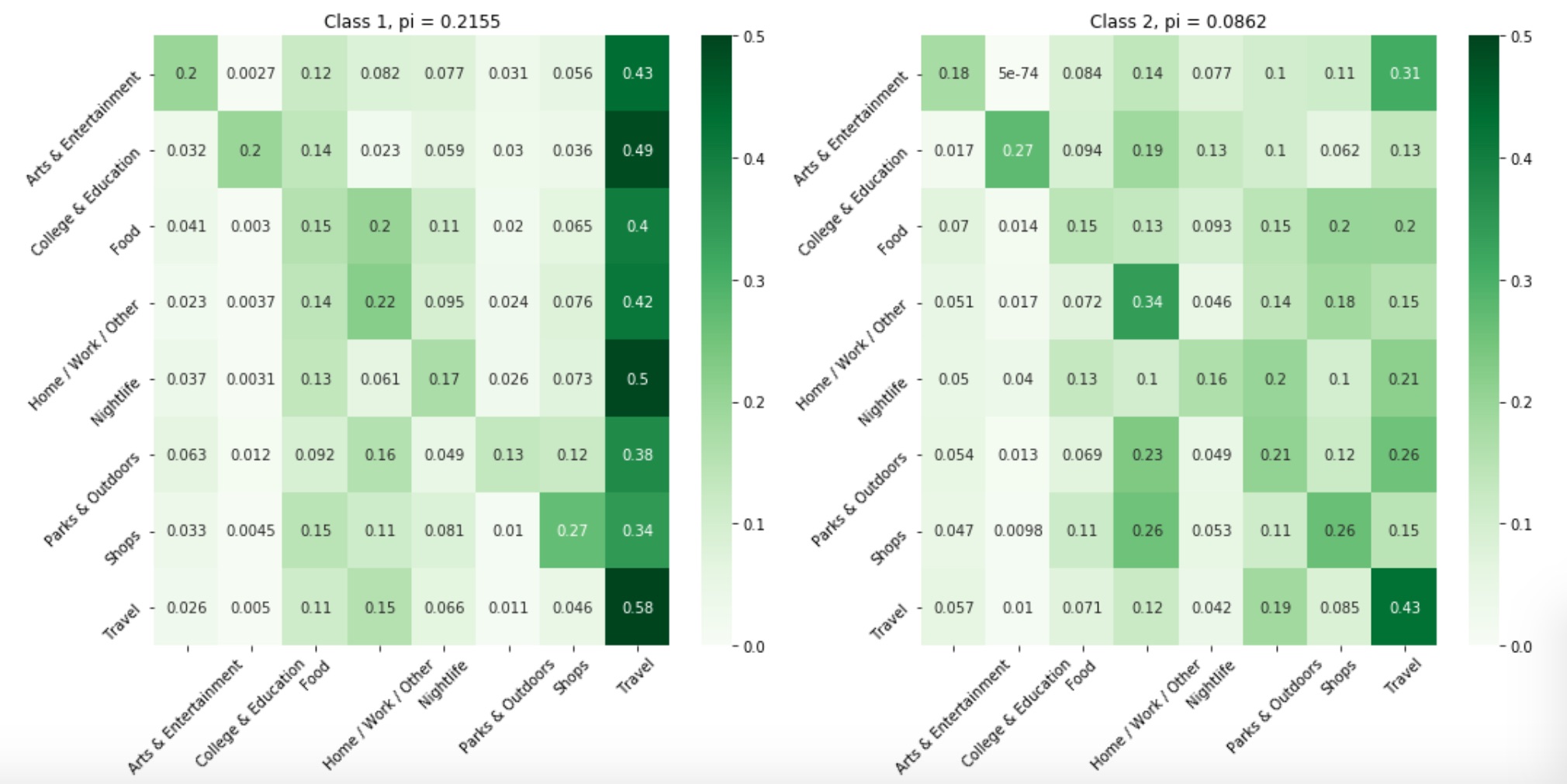}
\caption{London. K=4. Transition probability matrices for cluster 1  and cluster 2.}\label{sample image 1}
\end{figure}

\begin{figure}
\centering
\includegraphics[width=1\textwidth]{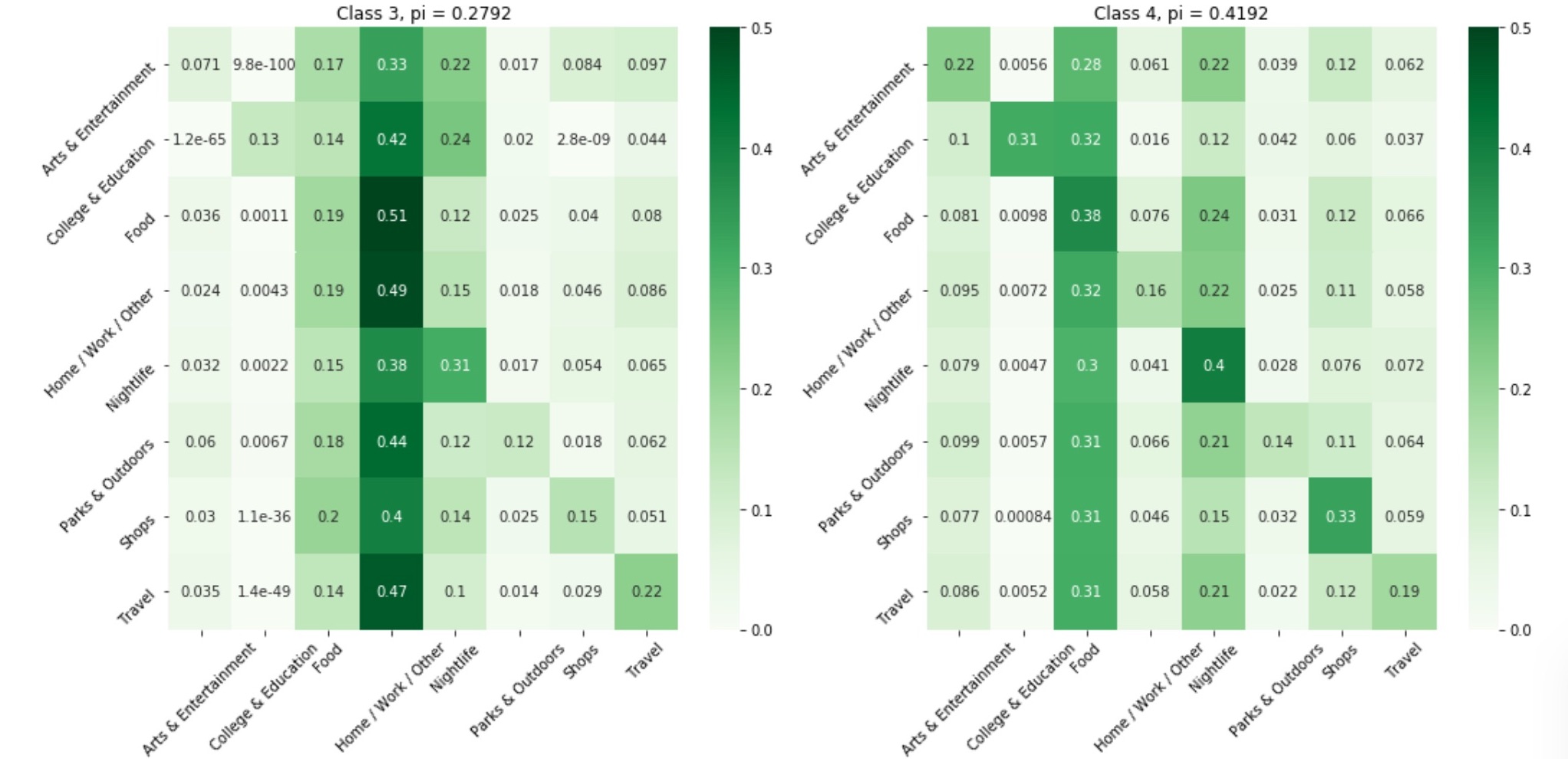}
\caption{London.  K=4. Transition probability matrices for cluster 3 and cluster 4.}\label{sample image 2}
\end{figure}

This means that the most popular transitions  in the biggest Cluster 4 are: Food - Food, Nightlife - Nightlife, Home/Work -   Food, Shops - Shops, and College $\&$ Education - Food. 

In Cluster 3,  the most popular transitions are:  Travel - Home/Work, Food - Home/Work, College $\&$ Education - Home/Work, and Home/Work - Home/Work.

In Cluster 1, the most popular transitions are: Travel - Travel, Nightlife - Travel, Art $\&$ Entertainment - Travel,  and College $\&$ Education - Travel.

In the smallest Cluster 2, the most popular transitions are: Travel -  Travel, Art $\&$ Entertainment  - Travel, Home/Work - Home/Work, Shops - Shops, and College $\&$ Education  - College $\&$ Education.

For each cluster, we also counted the sum of elements by columns of the corresponding transition probability matrix, and constructed the vector $sum_i=(sum_i^1, ..., sum_i^C)$  as follows:
$$sum_i^k= \sum_{j=1}^C T_i(j,k),\ k=1..C.$$ 
The quantity $sum_i^k$ shows how often the category $k$ becomes the next step in the sequence from  cluster $i=1,..,K$.  Correspondingly, the vector $sum_i$ shows the popularity of each category between users from  cluster $i$.  


For simplicity, in each cluster we consider only  3  of the most popular categories. This helps us to understand the peculiarities of each cluster, and also makes it possible to compare  similar clusters of various cities.

We did this analysis for  5 cities:
London,  Chicago, San Francisco, New York, and Brooklyn (for Brooklyn there are enough of check-ins to study this  district separately of New York). In all these cities, the number of clusters is 4.

In Table~\ref{tab:Table 1}, we compared the clustering of  
New York  and San Francisco. The mixing distribution (proportional to the cluster sizes) for New York  is  $$p^{users}_{NY}=(0.59, 0.15, 0.06, 0.20),$$ for San Francisco  is $$p^{users}_{SF}=(0.52, 0.27, 0.04, 0.17).$$ 

\begin{table}[htbp]
\footnotesize
  \caption{Comparison of clusters of New York and San Francisco.}
   \label{tab:Table 1}
   \begin{center}
   \begin{tabular}{ccc}
     \hline
Cluster& New York & San Francisco \\ 
i&  category j,\ \  popularity $sum_i^j$ & category j,\ \  popularity $sum_i^j$\\
     \hline
  & Food, 3.57 & Food, 3.68  \\ 
 1 & Nightlife, 1.70 & Nightlife, 1.65 \\     
 & Shops, 0.76 & Shops, 0.81  \\ 
  \hline
 & Travel,  2.17& Travel, 2.32\\ 
 2 & Food,  1.81& Food, 2.18\\ 
& Home/Work,  1.20& Shops , 0.99 \\ 
 \hline
 & Food, 1.94& Food, 2.08  \\
 3 & Home/Work, 1.22& Parks \& Outdoors, 1.95\\ 
 & Parks \& Outdoors, 1.19 & Shops, 1.33 \\ 
  \hline
 & Home/Work, 2.91& Home/Work, 2.67 \\ 
 4 & Food, 2.19& Food, 2.06\\ 
& Nightlife, 1.03 & Nightlife, 0.87 \\ 
\hline
\end{tabular}
\end{center}
\end{table}


Each cluster in these cities is represented by the  3 most popular categories. We also give the value of 
 $sum_i^j$  for these categories in these clusters. 

Additional remarks:
\begin{itemize}
\item In Brooklyn,  there is  a cluster, where the  Art $\&$ Entertainment - Art $\&$ Entertainment transition is one of the most popular. This phenomenon is discovered only in this place.
\end{itemize}

\begin{figure}
\centering
\includegraphics[width=1\textwidth]{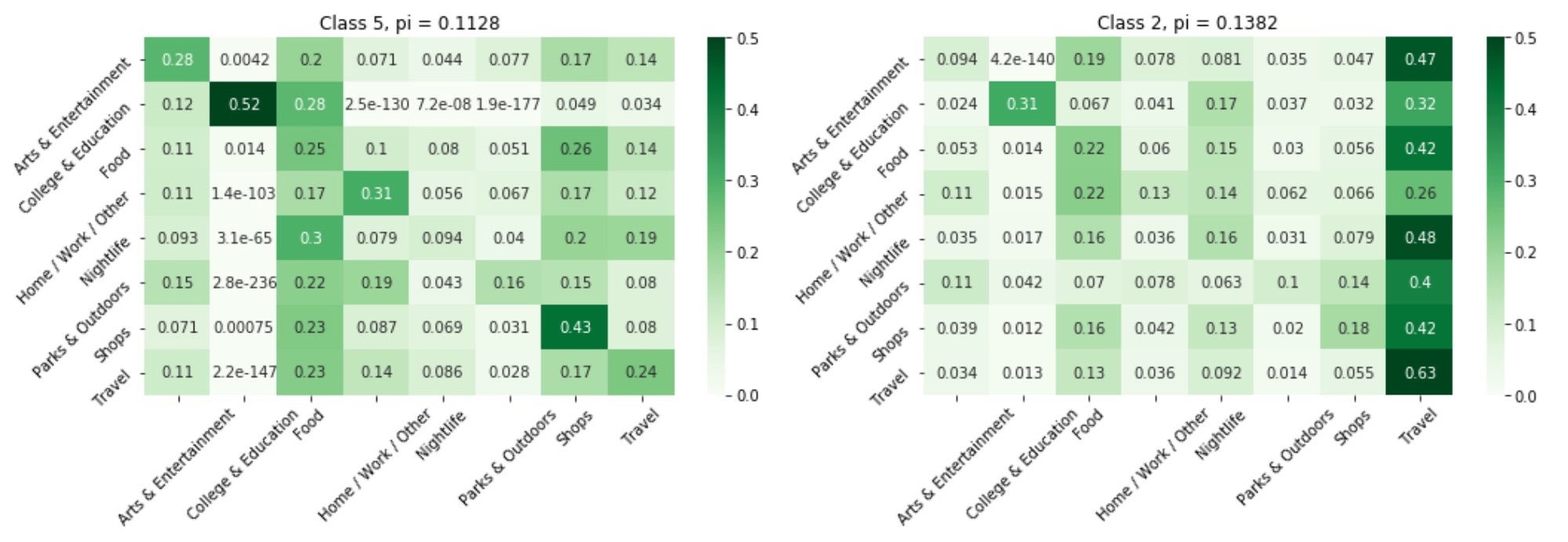}
\caption{London. K=6. Transition probability matrices for  cluster 2 and cluster 5.}\label{sample image 4}
\end{figure}

We also studied in more detail clustering for London using  6 clusters,
and noticed that the clusters became  more interpretable in this case, see Figure~\ref{sample image 4} for example.


\section{Discussion}
The method of our study gives a rather natural cluster division: 
in each city we could see a cluster with the behavior of party people (cluster 4 for London), a cluster for students (cluster 3 for London), a cluster for business people (cluster 1 for London), and the smallest cluster of people with nontypical behavior (cluster 2 for London, K=4).

We observe a similar picture for each city/place that we considered (London,  San Francisco, Chicago, New York, Brooklyn). The differences are in the sizes of these clusters and with certain peculiarities in several cities.

We can compare our results with the results of M. Gonzalez et al.,   \citep[see][]{gonzalez'18},  which  they obtained for Mexico City by analyzing  sequences of bank transactions. 
They had  6 clusters with different behavior: commuter (13\%), household (17\%), youth (11\%), hi-tech (10\%), average (39\%), and dinner-out (10\%).

We noticed several similarities with our results: we also have a cluster covering hi-tech people (business people cluster) and a cluster that corresponds to the dinner-out people (party people cluster). Nevertheless, we could see differences in our results.  By analyzing check-ins, we are able to observe only a certain subpopulation of citizens, namely active users of  geosocial networks. Thus, our sample could not be  a representative sample of the population of all citizens,
but the  advantage of our data is that they are easy to obtain and could be mined in real time. The data from bank transactions are less freely available and often private.

We  would also like to state that the natural and traceable results of our work  confirm that the mixture Markov chain  process (in combination with the EM algorithm) is a good, but underestimated tool for the clustering of the sequences of activities.

We would also like to mention the convenience of this model, namely that the parameters of Markov chains already give a lot of information about the clusters. This situation is different with other clustering methods. They need many additional computations for the  cluster's interpretation.

We would like to highlight the open problems of our study. We need to additionally  analyze the geographical component of the data to get more information for a description of the clusters that we obtained, also it would be good to adjust the available friendship information.

\section{Conclusions}
We clusterized  the check-in sequences  from the  geosocial network, Weeplaces, by using  the mixture Markov chain model.  This allowed us to obtain a good and traceable division of user behavior into 4 or 6  clusters.  Our next goals are to use geolocations for a better interpretation of  the clusters and the application of these methods to recent data from our region.

\section{Acknowledgements}
We are grateful to prof. D. Alexandrov for a very valuable discussion on the subject, also to N.Smirnova (the professor of the course on "Basics of writing an empirical research article") for her support during this paper writing.
\\
{\it{This research did not receive any specific grant from funding agencies in the public, commercial, or not-for-profit sectors.}}

\bibliography{mybibfile}{}

\end{document}